\documentclass[aps,prl,twocolumn,superscriptaddress,showpacs]{revtex4}
\usepackage{CJK}
\usepackage[Symbol]{upgreek}
\usepackage{bm}
\usepackage{amsmath}
\usepackage{graphicx}
\usepackage{mathptmx} 
\usepackage{endnotes}
\usepackage{color}
\usepackage{float}
\usepackage[FIGTOPCAP, hang, tight ]{subfigure}
\begin{document}
\begin{CJK*}{GB}{song}


\title{Prediction of Anisotropic Single-Dirac-Cones in Bi${}_{1-x}$Sb${}_{x}$ Thin Films  }


\author{Shuang Tang}
\email[]{tangs@mit.edu}
\affiliation{Department of Materials Science and Engineering, Massachusetts Institute of Technology, Cambridge, MA
02139-4037, USA} 
\author{Mildred S. Dresselhaus}
\email[]{millie@mgm.mit.edu}
\affiliation{Department of Electrical Engineering and Computer
Science, Massachusetts Institute of Technology, Cambridge, MA
02139-4037, USA} \affiliation{Department of Physics,Massachusetts
Institute of Technology, Cambridge, MA 02139-4037, USA}


\date{\today}

\begin{abstract}
The electronic band structures of Bi${}_{1-x}$Sb${}_{x}$ thin films can be varied as a function of temperature, pressure, stoichiometry, film thickness and growth orientation. We here show how different anisotropic single-Dirac-cones can be constructed in a Bi${}_{1-x}$Sb${}_{x}$ thin film for different applications or research purposes. For predicting anisotropic single-Dirac-cones, we have developed an iterative-two-dimensional-two-band model to get a consistent inverse-effective-mass-tensor and band-gap, which can be used in a general two-dimensional system that has a non-parabolic dispersion relation as in a Bi${}_{1-x}$Sb${}_{x}$ thin film system. 
\end{abstract}
\maketitle
\end{CJK*}
Materials with two-dimensional (2D) Dirac cones in their electronic band structures have recently attracted considerable attention. Many important novel physical studies have been carried out on both massless and massive 2D Dirac fermions, including studies of the half-integer quantum Hall effect \cite{1}, the anomalous absence of back scattering, the Klein paradox effect \cite{3}, high temperature superconductivity \cite{5} and unusual microwave response effects \cite{6}. The ultrahigh carrier mobility of the Dirac fermions in graphene offers new opportunities for a variety of electronics applications \cite{4}. Recently, 2D Dirac cones observed in topological insulators have identified this class of materials as candidates for quantum computation, spintronics, novel superconductors and thermoelectrics \cite{7}.

Materials with 2D single-Dirac-cones, especially 2D anisotropic single-Dirac-cones, are of special interest. Simulations with ultracold atoms trapped on optical lattices have been used to study general 2D single-Dirac-cones \cite{14}. Graphene superlattice materials, are described by anisotropic 2D single-Dirac-cones, which could potentially be developed for use in nano-electronic-circuits without cutting processes, and in table-top experiments that simulate high-energy relativistic particles propagating in anisotropic space \cite{15}.

Bi${}_{1-x}$Sb${}_{x}$ alloys could be especially attractive for studies of 2D Dirac fermions. Firstly, the band structure of three-dimensional (3D) bulk Bi${}_{1-x}$Sb${}_{x}$ can be varied as a function of Sb composition \textit{x}, temperature, pressure and strain, and the Fermi level can be adjusted to change the electronic properties \cite{16}. Secondly, not only have bulk state Dirac points been studied in bulk Bi${}_{1-x}$Sb${}_{x}$ alloys \cite{27}, but also the first observation of a surface state 2D single-Dirac-cone for a topological insulator was made in bulk Bi${}_{0.9}$Sb${}_{0.1}$ \cite{28}. Some experiments on Bi${}_{1-x}$Sb${}_{x}$ thin films grown normal to the trigonal direction have already been carried out \cite{39,40,41}. However, possible Dirac-cone materials of 2D Bi${}_{1-x}$Sb${}_{x}$ thin films have not yet been studied. 

In this Letter, we have systematically studied how the properties of the 2D \textit{L}-point Dirac cones in Bi${}_{1-x}$Sb${}_{x}$ thin films, such as their anisotropy and linearity, change with film thickness and growth orientation. We have also discovered that different varieties of Dirac cones can be constructed in Bi${}_{1-x}$Sb${}_{x}$ thin films, such as single-Dirac-cones with different anisotropies. For the methodology, we have developed an iterative-two-dimensional-two-band model, which could be used as a general model to study systems with non-parabolic 2D dispersion relations or Dirac cones. 
 
We recall that the band-gap and band-overlap of bulk Bi can be changed by adding Sb to form Bi${}_{1-x}$Sb${}_{x}$ alloys. 3D Dirac points may form under proper conditions at the three \textit{L} points \cite{27}, where the valence band and conduction band exhibit a band-crossing. At each \textit{L} point, $E(\mathbf{k})$ becomes linear in the limit of band-crossing ($E_{g}\to{0}$), since $E(\mathbf{k})=\pm \left((\mathbf{v}\cdot \mathbf{p})^{2} +E_{g}^{2} \right)^{\frac{1}{2} }$, where $\mathbf{v}$ is the carrier velocity, $\mathbf{p}$ is the carrier momentum, and $E_{g}$ is the \textit{L}-point direct band-gap \cite{28, 29}. Historically, the 3D two-band model \cite{30} has been successful in describing the non-parabolic dispersion relation $E(\mathbf{k})$ in 3D bulk Bi materials by 
\begin{equation} \label{GrindEQ__1_} 
\mathbf{p}\cdot \bm\upalpha \cdot \mathbf{p}=E(\mathbf{k})(1+\frac{E(\mathbf{k})}{E_{g} } ),                                                    
\end{equation} 
where $\mathbf{\bm\upalpha}$ is the inverse-effective-mass-tensor, and we assume here that $\bm\upalpha$ is the same for both the conduction band and valence band within the context of a two-band model and strong interband coupling. Generally, the relation between $\bm\upalpha$ and $E_{g}$ around an \textit{L} point is described as
\begin{equation} \label{GrindEQ__2_} 
\bm\upalpha =\frac{2}{\hbar ^{2} } \frac{\partial ^{2} E(\mathbf{k})}{\partial (\mathbf{k}-\mathbf{k}_{L} )^{2} } 
=\frac{1}{m_{0} } \cdot \mathbf{I}\pm \frac{1}{m_{0}^{2} } \frac{2}{E_{g} } \cdot \mathbf{p}^{2} ,                                    
\end{equation} 
under the $\mathbf{k}\cdot \mathbf{p}$ approximation \cite{31}, where $\mathbf{I}$ is the identity matrix and $m_{0}$ is the free electron mass. We assume here that a two-band model also applies to Bi${}_{1-x}$Sb${}_{x}$ alloys, where the influence on $\bm\upalpha $ of adding Sb atoms up to a Sb concentration of \textit{x}=0.1 to bulk Bi \cite{31} is then
\begin{equation} 
\label{GrindEQ__3_} 
\bm\upalpha (Bi_{1-x} Sb_{x} )=\frac{E_{g} (Bi)}{E_{g} (Bi_{1-x} Sb_{x})} \cdot (\bm\upalpha 
(Bi)-\frac{1}{m_{0}} \cdot\mathbf{I})+\frac{1}{m_{0}}\cdot 
\mathbf{I}.                    
\end{equation}

We developed an iterative-two-dimensional-two-band model for Bi${}_{1-x}$Sb${}_{x}$ thin films, where both $\bm\upalpha ^{film} $ and $E_{g}^{film} $ of a Bi${}_{1-x}$Sb${}_{x}$ thin film differ from $\bm\upalpha ^{bulk} $ and $E_{g}^{bulk}$ in the bulk case, and they are both unknown at the start. The conduction band and valence band at an \textit{L} point are firstly separated by $E_{g}^{[0]} =E_{g}^{bulk}(Bi)$ and the inverse-effective-mass-tensor is $\bm\upalpha ^{[0]} =\bm\upalpha ^{bulk} (Bi)$. In a Bi${}_{1-x}$Sb${}_{x}$ thin film, the \textit{L}-point gap will be influenced by both the added Sb atoms and the quantum confinement effect, so that the \textit{L}-point gap to lowest order becomes
\begin{equation}
\label{GrindEQ__4_}
 E_{g}^{[1]} =E_{g}^{bulk} (Bi_{1-x} Sb_{x} )+2\cdot \frac{h^{2} \upalpha_{33}^{[0]} }{8\cdot l_{z}^{2} } . 
\end{equation}
The change in the \textit{L}-point gap will lead to a change of the \textit{L}-point inverse-effective-mass-tensor according to
\begin{equation} 
\label{GrindEQ__5_} 
\bm\upalpha ^{[n]} =\frac{E_{_{g} }^{[n-1]} }{E_{_{g} }^{[n]} } \cdot \bm\upalpha ^{[n-1]} 
+\frac{1}{m_{0} } \cdot (1-\frac{E_{_{g} }^{[n-1]} }{E_{_{g} }^{[n]} } )\cdot \mathbf{I},
\end{equation} 
where n denotes the step number in the iteration. Eq. (5) is a general equation for iterations, which is just a consequence of Eq. (3). Thereafter, the \textit{L}-point band-gap can be updated with the new inverse-effective-mass-tensor by 
\begin{equation} 
\label{GrindEQ__6_}
E_{g}^{[n+1]} =E_{g}^{[n]}+2\cdot \frac{h^{2} \upalpha_{33}^{[n]} }{8\cdot l_{z}^{2} }.
\end{equation} 
The procedure is repeated until $\bm\upalpha ^{[n]} $ and $E_{g}^{[n]} $ become self-consistent, and then we get an accurate solutions for $\bm\upalpha^{[n]} =\bm\upalpha ^{film} (Bi_{1-x} Sb_{x} )$ and $E_{g}^{[n]} =E_{g}^{film} (Bi_{1-x} Sb_{x} )$ for the Bi${}_{1-x}$Sb${}_{x}$ thin film. The Hamiltonian for Bi and Bi${}_{1-x}$Sb${}_{x}$ based on $\mathbf{k}\cdot \mathbf{p}$ theory in this model is equivalent to a Dirac Hamiltonian with a scaled canonical conjugate momentum \cite{29}. Thus, Eqs. (5) and (6) are also good approximations to describe the Dirac cones. The band parameters we use in the present calculations are values that were measured by cyclotron resonance experiments \cite{36}.

\begin{figure}
        \includegraphics[width=0.45\textwidth]{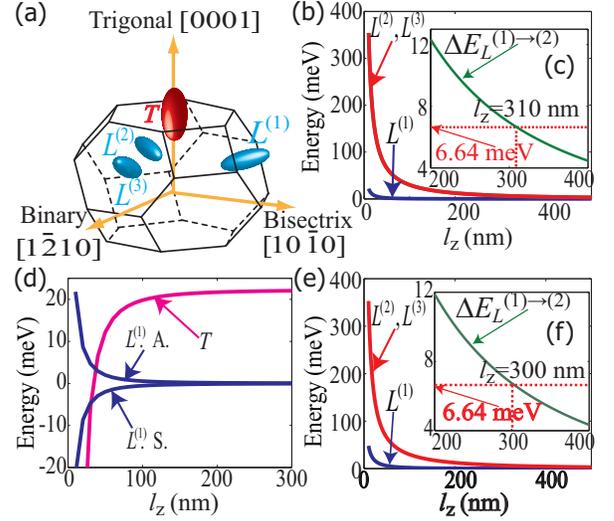}
        \caption{(color online).     Schematics of the conduction band and valence band for different carrier pockets of a Bi${}_{0.96}$Sb${}_{0.04}$ thin film vs. film thickness: (a) illustrates the \textit{T}-point hole pocket and the three \textit{L}-point electron pockets in the first Brillouin zone for bulk bismuth with a rhombohedral crystal structure. (b)-(d) are for Bi${}_{0.96}$Sb${}_{0.04}$ thin films that are grown normal to the bisectrix direction associated with the $L^{(1)}$ point. E=0 corresponds to the middle point of the \textit{L}-point band-gap. (b) and (c) are comparisons between energies of the conduction band at the three \textit{L} points, $L^{(1)}$, $L^{(2)}$ and $L^{(3)}$ for different film thicknesses. $\Delta E_{L}^{(1)\to (2)}$ is smaller than the thermal smearing the energy of a carrier at $L^{(1)}$ at 77 K which corresponds to $l_{z} \le$ 310nm from the iteration of Eqs. (4), (5) and (6). (d) Comparison between energies of the conduction band and valence band at the $L^{(1)}$ point and the valence band at the \textit{T} point. (e) and (f) are comparison between the energies of the conduction band at the three \textit{L} points, $L^{(1)}$, $L^{(2)}$ and $L^{(3)}$, for different film thicknesses $l_{z}$  when the Bi${}_{0.96}$Sb${}_{0.04}$ thin film is grown normal to the ${\rm [60}\bar{{\rm 6}}{\rm 1]}$ direction. }
\end{figure}

In order to get a single-Dirac-cone, where only one Dirac cone effectively contributes Dirac carriers in the 2D system, we need to grow the Bi${}_{1-x}$Sb${}_{x}$ thin film normal to a low symmetry direction so that we can have a single \textit{L}-point that differs from the other two by breaking the 3-fold symmetry occurring in 3D. Moreover, two quantities in Eq. (4) need to be minimized, namely the bulk-band-gap term $E_{g}^{bulk}(Bi_{1-x} Sb_{x})$ and the quantum-confinement-induced term ${h^{2} \upalpha_{33}^{[0]} }/{4\cdot l_{z}^{2}}$ for this special single \textit{L}-point. Thus, by growing a film normal to the bisectrix direction the 3-fold symmetry is broken and the $\upalpha_{33}$ inverse mass component is near its minimum \cite{36}, which is discussed below. Minimizing the value of bulk term $E_{g}^{bulk}(Bi_{1-x} Sb_{x})$ by varying temperature T, pressure P and Sb composition \textit{x} has been discussed in literature \cite{28,31,34}, especially for the case of T=77 K P=1 atm and \textit{x}=0.04, where $E_{g}^{bulk}(Bi_{1-x} Sb_{x})$ ceases to 0. Bi${}_{1-x}$Sb${}_{x}$ thin films with this composition of \textit{x}=0.04 (Bi${}_{0.96}$Sb${}_{0.04}$) have been already synthesized in experiments by  researchers in Ukraine \cite{27} for films grown normal to the trigonal direction. In the present paper, we illustrate simple examples of the various kinds of single-Dirac-cones that can be found in Bi${}_{0.96}$Sb${}_{0.04}$ for more generally and in particular for conditions where Dirac cones can be found. 

According to our calculations, a single 2D Dirac cone can be constructed 
in a Bi${}_{0.96}$Sb${}_{0.04}$ thin film by breaking the 3-fold symmetry of the \textit{L} points. We calculate here the lowest conduction band and the highest valence band at the three $L$ points, namely $L^{(1)}$, $L^{(2)}$ and $L^{(3)}$ shown in Fig. 1(a), when the Bi${}_{0.96}$Sb${}_{0.04}$ thin film is grown normal to the bisectrix direction associated with the $L^{(1)}$ point [Fig. 1(b)] in this case. The $L^{(2)}$ and $L^{(3)}$ points still retain a double degeneracy with each other, while the energy difference between the lowest conduction band at the $L^{(1)}$ point and the corresponding energy at the $L^{(2)}$ point or $L^{(3)}$ point ($\Delta E_{L}^{(1)\to (2)}=E_{L}^{(2)}-E_{L}^{(1)}$, $\Delta E_{L}^{(1)\to (2)} $=$\Delta E_{L}^{(1)\to(3)}$) increases as the film thickness decreases [Fig. 1(c)]. Below 77 K the band structure of Bi${}_{1-x}$Sb${}_{x}$ does not vary significantly with temperature.  At 77 K, the smearing of the Fermi-Dirac distribution is $\sim k_{B} T$=6.64 meV. Thus, when $\Delta E_{L}^{(1)\to (2)} \ge 6.64$ meV, the Dirac fermions can all be associated with the $L^{(1)}$ point. The energy difference $\Delta E_{L}^{(1)\to(2)}=6.64$ meV corresponds to a film thickness of $l_{z}=310$ nm as shown in Fig. 1(c). Thus, we can have a 2D single-Dirac-cone at the $L^{(1)}$ point in a Bi${}_{0.96}$Sb${}_{0.04}$ film that is grown normal to the bisectrix direction and thinner than 310 nm.

In experiments, a Dirac cone is usually not perfect. A mini-gap often 
exists which induces a tiny mass at the apex of the Dirac cone. Such an effect also occurs in single layer graphene \cite{1}. There are two main features that characterize the properties of a Dirac cone, the tiny mass occuring at the apex of the Dirac cone and the fermion velocity $\mathbf{v}(\mathbf{k})$ as a function of $\mathbf{k}$. Because the fermions are linearly dispersed near a Dirac cone, $\mathbf{v}(\mathbf{k})$ should only be a function of the direction of $\mathbf(k)$. The anisotropy of the Dirac cone can be characterized by the ratio between the maximum and minimum $\mathbf{v}(\mathbf{k})$. 

\begin{figure}
        \includegraphics[width=0.5\textwidth]{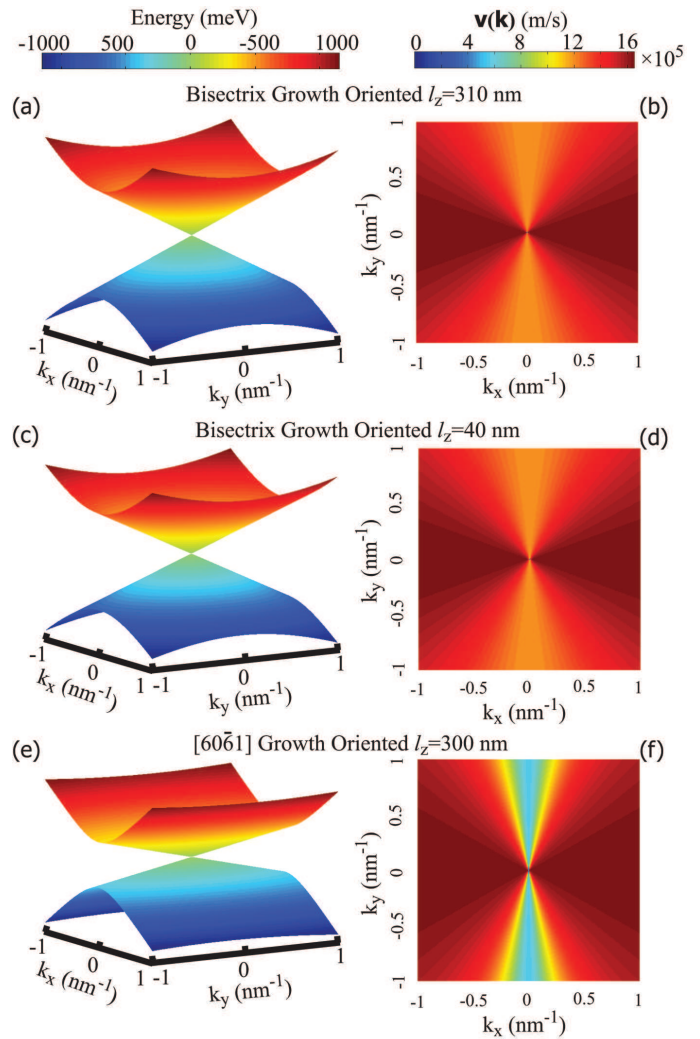}
        \caption{(color online).     Different anisotropic single-Dirac-cones in different Bi${}_{0.96}$Sb${}_{0.04}$ thin films for various applications or research purposes: (a) and (b) describe a sharp-apex anisotropic single-Dirac-cone in a 310 nm thick Bi${}_{0.96}$Sb${}_{0.04}$ film grown normal to the bisectrix direction. For convenience, the origin of momentum $\mathbf{k}$ is chosen to be at the $L^{(1)}$ point.  (c) and (d) describe an anisotropic single-Dirac-cone where the Dirac electrons are free from the influence of \textit{T}-point holes in a 40 nm thick Bi${}_{0.96}$Sb${}_{0.04}$ film grown normal to the bisectrix direction. (e) and (f) describe a highly anisotropic single-Dirac-cone in a 300 nm thick Bi${}_{0.96}$Sb${}_{0.04}$ film grown normal to the ${\rm [60}\bar{{\rm 6}}{\rm 1]}$ direction. (a), (c) and (e) show the dispersion relation of single-Dirac-cones. (b), (d) and (f) show the group velocity $\mathbf{v}$ of Dirac fermions at different values of momentum $\mathbf{k}$. (c)-(d) are not significantly different from (a)-(b), and (e)-(f) are obviously more anisotropic than (a)-(b) and (c)-(d). }
\end{figure}

The Dirac cone at the $L^{(1)} $ point in a 310 nm thick Bi${}_{1-x}$Sb${}_{x}$ thin film grown normal to the bisectrix axis has a linear $E(\mathbf{k})$ behavior with a very sharp apex [Fig. 2(a)]. In this film, k${}_{x}$ and k${}_{y}$ are in the trigonal direction and the binary direction, respectively. $E_{g}^{(1)} $ for this Dirac cone is smaller than 0.1 meV, and the effective mass at the apex of the Dirac cone is $\sim 10^{-5} m_{0} $, which can be considered as essentially massless. We have also calculated the $\mathbf{v}(\mathbf{k})$ relation of the Dirac fermions for different values of momentum [Fig. 2(b)]. It can be seen that the maximum and minimum of $\mathbf{v}(\mathbf{k})$ are $1.6\cdot c_{0}/300$ (along k${}_{x}$) and $1.1\cdot c_{0}/300$, (along k${}_{y}$) respectively, which differs by a  factor of ${\sim}1.5$, where $c_{0}$ is the speed of light.  

In some applications, we need all the carriers within the film to 
be linearly dispersed, which also means that the Fermi level ${E}_{f}$ needs to be arranged to have no \textit{T}-point parabolically dispersed carriers. This condition is not met in bulk Bi${}_{0.96}$Sb${}_{0.04}$ , but it can be achieved in a Bi${}_{0.96}$Sb${}_{0.04}$ thin film. The highest valence band at the \textit{T} point by comparison to that of the lowest conduction band at the $L^{(1)} $ point is calculated in Fig. 1(d). Based on Fig. 1(d), when $l_{z}\le 40$ nm the film becomes an indirect gap semiconductor and if ${E}_{f}$ is adjusted to be high enough above the top of the valence band, carriers can all be $L^{(1)}$-point electrons, with no \textit{T}-point holes. At this film thickness, $E_{g}^{(1)} $ at the $L^{(1)} $ point is smaller than 4 meV, and the effective mass at the apex of the Dirac cone is $\sim 10^{-4}{}m_{0} $, which is still essentially massless.  The corresponding Dirac cone is plotted [Fig. 2(c)], as well as the velocity vs. momentum relation $\mathbf{v}(\mathbf{k})$ for the Dirac fermions [Fig. 2(d)]. The linearity and the anisotropy of the Dirac cone is not notably influenced by film thickness for a film of $l_{z} =40$ nm, as can be seen by comparing Figs. 2(c) and 2(d) to Figs. 2(a) and 2(b), respectively. When $l_{z} \le 31$ nm, the film become a direct gap semiconductor [Fig. 1(d)], and the both the Dirac electrons and holes can be free from the influence of the \textit{T}-point 
parabolically dispersed holes if the Fermi level is above the top of the valence band. 

In some other applications, a higher anisotropy of the Dirac cone 
could be required, e.g. in nano-electronic-circuit design. $E(\mathbf{k})$ at the $L^{(1)} $ point of a Bi${}_{0.96}$Sb${}_{0.04}$ thin film grown normal to the corresponding bisectrix direction has a smaller tiny-apex-mass but a lower anisotropy, while $E(\mathbf{k})$ at the $L^{(1)}$ point of a Bi${}_{0.96}$Sb${}_{0.04}$ thin film grown normal to the trigonal direction has 
a larger tiny-apex-mass but a higher anisotropy, when compared to Bi${}_{0.96}$Sb${}_{0.04}$ thin films of the same thickness. To get a single-Dirac-cone with both a small tiny-apex-mass and high anisotropy, we can choose to grow the film normal to a direction between the trigonal $[0001]$ axis and the bisectrix axis $[10\bar{1}0]$ within the binary plane, where $[0001]$ and $[10\bar{1}0]$ are hexagonal 
notations for such crystalline directions in bulk rhombohedral Bi${}_{1-x}$Sb${}_{x}$. As an example, we illustrate in Fig. 2(e) and (f) the film grown normal to the ${\rm [60}\bar{{\rm 6}}{\rm 1]}$ direction, which is a lower-symmetry direction in the binary plane with an angle of $14^{\circ}$ to the bisectrix axis and $76^{\circ}$ to the trigonal axis. Results for the lowest conduction band and the highest valence band at the three $\mathit{L}$ points as calculated for a Bi${}_{0.96}$Sb${}_{0.04}$ thin film grown normal to the ${\rm [60}\bar{{\rm 6}}{\rm 1]}$ direction are shown in Fig. 1(e). According to the criterion we used above [Fig. 1(b)], the energy dependance between $E_{L}^{(1)}$ and $E_{L}^{(2)}$ is $\Delta E_{L}^{(1)\to(2)} \le 6.64$ meV when $l_{z} \le 300$ nm [Fig. 1(f)]. The $L^{(1)}$ point Dirac cone in the 300 nm thick Bi${}_{0.96}$Sb${}_{0.04}$ thin film grown normal to the ${\rm [60}\bar{{\rm 6}}{\rm 1]}$ direction is plotted [Fig. 2(e)]. In this film, k${}_{y}$ is still normal to the binary direction, and k${}_{x} $ is in the direction perpendicular to both the binary direction and the ${\rm [60}\bar{{\rm 6}}{\rm 1]}$ direction. $E_{g}^{(1)} $ for this Dirac cone [Fig. 2(e)] is smaller than 0.46 meV, and the effective mass at the apex of the Dirac cone is still negligible $(\sim 10^{-4} m_{0})$. The $\mathbf{v}(\mathbf{k})$ relation of the Dirac fermions calculated for this cone is shown in Fig. 2(f). The linearity and the sharpness of $E(\mathbf{k})$ are not sacrificed significantly compared to Fig. 2(a). However, the anisotropy of the Dirac cone in Fig. 2(e) is increased remarkably. The maximum and minimum values of the Dirac fermion velocity $\mathbf{v}(\mathbf{k})$ near this Dirac cone are, respectively, $1.65\cdot c_{0}/300$ and $0.55\cdot c_{0}/300$, which differ by a factor of ${\sim}3$ [Fig. 2(f)], which we consider to be a highly anisotropic single-Dirac-cone.

The Fermi level can be adjusted by doping such a Bi${}_{1-x}$Sb${}_{x}$ sample with foreign elements, as for example by doping the sample with selenium or tellurium (Column VI) to increase the Fermi level (n-type) and by doping with tin or lead (Column IV) to decrease the Fermi level (p-type). In the graphene system, the Fermi level cannot be adjusted freely by heavy doping, because the band distortion due to heavy doping may destroy the linear \textbf{k} aspect of the Dirac cone itself. However, in a Bi${}_{1-x}$Sb${}_{x}$ thin film, the Sb composition, e.g. \textit{x}=0.04, is over 2 orders of magnitude larger than normal electronic doping concentrations ($\le 10^{-4} $), so that the band structures will not be obviously distorted by electronic doping.  Thus, the Fermi level can 
be varied over a wide range of doping levels relative to $E=0$ in a Bi${}_{1-x}$Sb${}_{x}$ thin film. 

In conclusion, we have developed an iterative-two-dimensional-two-band 
model to describe a general 2D non-parabolic anisotropic dispersion relation. Based on this methodology, we have made a prediction that anisotropic single-Dirac-cones can be constructed in Bi${}_{1-x}$Sb${}_{x}$ thin films. Some critical cases of $L^{(1)} $ point single-Dirac-cones are illustrated as examples: a single-Dirac-cone with a sharp apex [Fig. 2(a)], a single-Dirac-cone without influence of \textit{T}-point parabolically dispersed holes in a Bi${}_{1-x}$Sb${}_{x}$ thin film [Fig. 2(c)] and a single-Dirac-cone with high momentum anisotropy [Fig. 2(e)]. Novel physical phenomena associated with Dirac cones and Dirac fermions that have been previously reported in other materials systems could hopefully also be observed in Bi${}_{1-x}$Sb${}_{x}$ thin films. Because the Bi${}_{1-x}$Sb${}_{x}$ thin film system has special features as we discussed above, we can also expect to observe new physical phenomena that have never been observed in other systems.

\begin{acknowledgments}
The authors acknowledge E. Rogacheva, J. Heremans, O. Rabin and A. Levin
for discussions and support from AFOSR MURI Grant number FA9550-10-1-0533, subaward 60028687. The views expressed are not endorsed by the
sponsor.
\end{acknowledgments}

\end{document}